\documentclass[journal=jacsat,manuscript=article]{achemso}

\usepackage[version=3]{mhchem} 

\usepackage{physics}
\usepackage{xargs}
\usepackage{booktabs}
\usepackage[dvipsnames]{xcolor}
\usepackage{chemformula} 
\usepackage{rotating} 
\usepackage{amsmath}

\usepackage[normalem]{ulem}
\usepackage{notes2bib}

\SectionNumbersOn 

\newcommand{\cfour}{\textsc{cfour }} 
\newcommand{\mint}{\textsc{mint }} 

\newcommand{\onlinecite}[1]{\hspace{-1 ex} \nocite{#1}\citenum{#1}} 

\author{Sophia Burger}
\affiliation{Department Chemie, Johannes Gutenberg-Universit\"at Mainz, Duesbergweg 10-14, D-55128 Mainz, Germany}
\email{soburger@uni-mainz.de}
\author{Stella Stopkowicz}
\email{stella.stopkowicz@uni-saarland.de}
\affiliation{Fachrichtung Chemie, Universität des Saarlandes, Campus B2.2, D-66123, Saarbr\"ucken, Germany}
\alsoaffiliation{Hylleraas Centre for Quantum Molecular Sciences, Department of Chemistry, University of Oslo, P.O. Box, Blindern, 1033, N-0315 
Oslo, Norway}
\author{J\"urgen Gauss}
\email{gauss@uni-mainz.de}
\affiliation{Department Chemie, Johannes Gutenberg-Universit\"at Mainz, Duesbergweg 10-14, D-55128 Mainz, Germany}

\title{
Cholesky Decomposition and the Second-Derivative Two-Electron Integrals Required for the Computation of Magnetizabilities using Gauge-Including Atomic Orbitals
}

\begin{document}


\begin{tocentry}





\includegraphics[width=\textwidth]{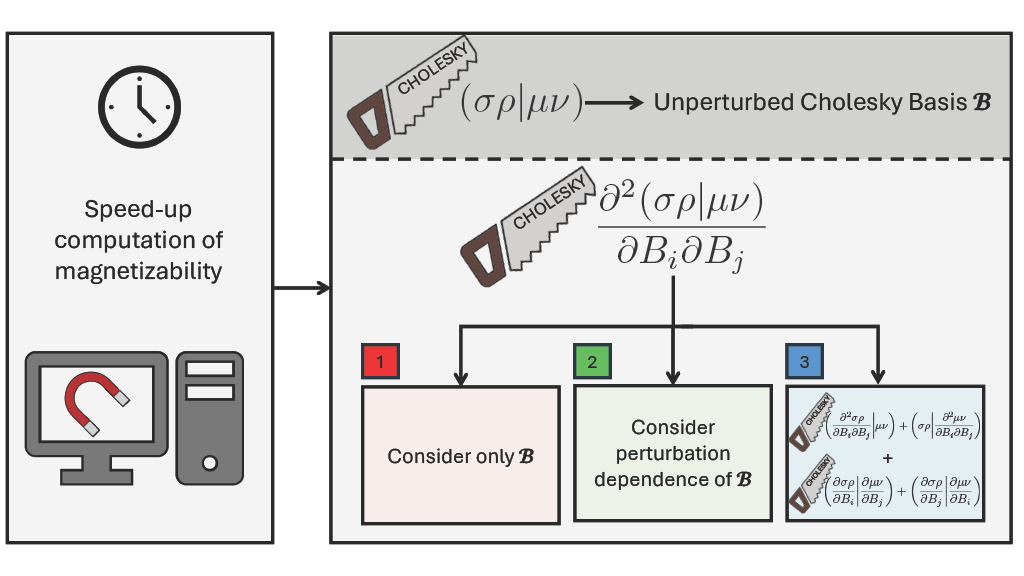}

\end{tocentry}


\begin{abstract}
The computation of magnetizability tensors using gauge-including atomic orbitals is discussed in the context of Cholesky decomposition for the two-electron
repulsion integrals with a focus on the involved doubly differentiated integrals. Three schemes for their handling are suggested: the first exploits the DF aspect of 
Cholesky decomposition, the second uses expressions obtained by differentiating the CD expression for the unperturbed two-electron integrals, while the third 
addresses the issue that the first two schemes are not able to represent the doubly differentiated integrals with arbitrary accuracy. This scheme uses a separate 
Cholesky decomposition for the cross terms in the doubly differentiated two-electron integrals. Test calculations reveal that all three schemes are able 
to represent the integrals with similar accuracy and yield indistinguishable results for the values of the computed magnetizability tensor elements. Thus, we recommend 
our first scheme which has the lowest computational cost for routine computations. The applicability of our CD schemes is further shown in large-scale Hartree-Fock
calculations of the magnetizability tensor of coronene (C$_{24}$H$_{12}$) with a doubly polarized triple-zeta basis consisting of 684 basis functions.
 
\end{abstract}


\section{Introduction}

Cholesky decomposition (CD) of the two-electron repulsion integrals has been proven an  efficient means for speeding up quantum-chemical computations, thus  enabling the quantum-chemical treatment of larger molecules.\cite{Pedersen24} CD of the two-electron integrals was first proposed by Beebe and Linderberg\cite{Beebe77} in the 1970ies. Koch  {\it et al.} have later demonstrated the great potential of this idea for quantum-chemical computations.\cite{Koch03} Today, CD is used to enable large-scale second-order M{\o}ller-Plesset (MP2) perturbation theory,\cite{Koch03,Aquilante07b,Bozkaya14,Blaschke22} multiconfigurational self-consistent-field,\cite{Aquilante08a,Nottoli21} multiconfigurational second-order perturbation theory,\cite{Aquilante08b} and coupled-cluster (CC) treatments.\cite{Pedersen04,Pitonak11,Epifanovsky13,Bozkaya16,Bozkaya20,Nottoli23,Zhang24,Uhlirova24,Blaschke24} The key idea of the CD of two-electron integrals is that the two-electron integrals as four-index quantities are represented as a product of three-index quantities, i.e., the so-called Cholesky vectors, which are obtained by applying a decomposition first proposed by Cholesky\cite{Benoit24} to the symmetric and semi-positive definite two-electron integral matrix. In this way, the cost for evaluating the two-electron integrals is reduced (as only a subset of them needs to be computed), I/O bottlenecks are avoided (as the resulting Cholesky vectors can typically be held in memory unlike the full set of two-electron integrals), and the resulting programs can be parallelized rather easily. In recent years, further improvements to the CD scheme have been suggested; noteworthy is here in particular the idea\cite{Aquilante11,Folkestad19,Zhang21,Gauss23} to replace the original one-step decomposition procedure by a two-step scheme, in which the Cholesky basis for the decomposition is determined in a first step and the actual Cholesky vectors are computed in the second step.

While CD is nowadays routinely used in quantum-chemical computations of energies and wave functions,\cite{Pedersen24} the use of CD in the computation of molecular properties is less explored. Schemes have been suggested for the computation of nuclear forces \cite{Aquilante08c,Feng19,Schnack22,Oswald24} required for geometry optimizations and the computation of nuclear magnetic resonance (NMR) shielding constants.\cite{Burger21,Nottoli22,Gauss23} The quantum-chemical calculation of both properties involve the computation of differentiated two-electron integrals. The challenge is, how one deals with the differentiated two-electron integrals in the CD context. CD cannot be applied directly to differentiated two-electron integrals, because their matrix representation does not constitute a symmetric/Hermitian semi-positive definite matrix. However, CD expressions for differentiated two-electron integrals can be obtained by differentiating the original CD expression for the undifferentiated two-electron integrals. It has been demonstrated that this idea works well for nuclear forces and NMR shielding constants and provides appropriate expressions for differentiated Cholesky vectors. Some reorganization of the obtained expressions allow computational improvements and in particular a further reduction of the computational cost.\cite{Gauss23,Schnack22}

The use of CD in computations that require second derivatives of the two-electron integrals, has so far not been investigated. The interesting question is here whether for those integrals suitable CD expressions can be obtained via differentiation. To a certain extent, this depends also on the objective: Do we want to obtain a good approximation to the results (integrals, values for properties) obtained in computation without CD  (in our opinion, what one should aim for) or yield just the correct second derivative of the energy within the CD procedure. In this paper, we explore the use of CD for the computation of the second derivatives of the two-electron integrals with respect to an external magnetic field. These derivative integrals are required for the computation of the magnetizability tensor\cite{Ruud93,Gauss07} when using gauge-including atomic orbitals (GIAOs).\cite{London37,Hameka58,Ditchfield72,Wolinski90,Helgaker91,Tellgren08} We discuss the challenges and offer solutions which enable treatments consistent with the overall CD procedure.

\section{Theory}\label{sec:theory}

CD allows to approximate the two-electron integrals as follows
\begin{eqnarray}
\label{eq1}
(\mu \nu |\sigma \rho) \approx \sum_P L_{\mu \nu}^P {L^P_{\rho \sigma}}^*
\end{eqnarray}
where the two-electron integrals $(\mu \nu | \sigma \rho)$ are denoted in Mulliken notation and Greek indices $\mu, \nu, \dots$ are used to label the atomic orbitals (AOs) $\chi_{\mu}, \chi_{\nu}, \dots$. The components of the $P$-th Cholesky vector (CV) are given by $L_{\mu\nu}^P$ and can be determined iteratively, as often described 
in the literature,\cite{Beebe77,Koch03} without computing the full two-electron integral matrix. Furthermore, a predefined Cholesky threshold $\tau$ is used to determine how many CVs are needed. This means that the iterative CD procedure is continued until the largest updated remaining diagonal is lower than the threshold $\tau$. The threshold further provides a rigorous error control, as the reconstructed integrals (via Eq. (\ref{eq1})) differ at most by $\tau$ from the actual integrals. We also note that the permutational symmetry of the two-electron integrals dictates that the CVs are symmetric with respect to an exchange of the two AO indices
\begin{eqnarray}
L_{\mu \nu}^P = L_{\nu \mu}^P
\end{eqnarray}
provided the used AOs are real what is assumed in the following.
It has further been shown that CD is closely related to density fitting (DF) in the way that Eq.~(\ref{eq1}) can be seen as DF with an on-the-fly determined orthogonalized auxiliary basis.\cite{Aquilante09a} This relationship between CD and DF turns out to be useful and will also invoked in the following. With ${\cal B} = \lbrace |P), \dots \rbrace$ as the Cholesky basis (CB) consisting of a set of orthonormal functions $|P)$, Eq.~(\ref{eq1}) can be rewritten as
\begin{eqnarray}
\label{eq2}
(\mu \nu |\sigma \rho) \approx \sum_{P\in {\cal B}}  (\mu \nu | P ) ( P | \sigma \rho)
\end{eqnarray}
with the CVs given as
\begin{eqnarray}
\label{eq3}
L_{\mu \nu}^P = (\mu \nu |P).
\end{eqnarray}
We note that the complete CB (the largest set that can be determined in the CD) is given by the full set of (orthonormalized) products of AO basis functions. This full set, denoted as ${\cal R}$, is larger than ${\cal B}$, but not equal to the complete basis, denoted in the following as ${\cal S}$. The latter also contains basis functions outside the used AO basis that cannot be represented by the products of the AOs. One can thus write the CD of the two-electron integrals as
\begin{eqnarray}
\label{eq4}
(\mu \nu |\sigma \rho) = \sum_{P\in {\cal B}}  (\mu \nu | P ) ( P | \sigma \rho) +  \sum_{P\in {\cal R  \backslash {\cal B}}}  (\mu \nu | P ) ( P | \sigma \rho) +  \sum_{P\in {\cal S} \backslash {\cal R}}  (\mu \nu | P ) ( P | \sigma \rho).
\end{eqnarray}
The first term in Eq.~(\ref{eq4}) is what is used for the integral when the CD approximation is invoked, the second term is what is neglected and assumed to be small (and shown to be smaller than $\tau$), while the third contribution is always exactly zero.  So far, we assumed that the
functions of the CB, in the following referred to as Cholesky basis functions, are orthonormal with respect to each other. However, it is often more useful to work with unorthogonalized basis functions which are simply given by the pairs of AO basis functions chosen in the CD procedure. Eq.~(\ref{eq1}) then reads as
\begin{eqnarray}
\label{eq5}
(\mu \nu |\sigma \rho) \approx \sum_{(\lambda, \chi), (\tau\xi) \in {\cal B}} (\mu \nu | \lambda\chi) (\lambda\chi | \tau \xi)^{-1} (\tau \xi |\sigma \rho)
\end{eqnarray} 
with the sum running over the chosen AO pairs.

Considering the permutational symmetry of the two-electron integrals, one might also use $1/2( \chi_{\mu}\chi_{\nu}^* + \chi_{\nu}\chi_{\mu}^*)$ instead of $\chi_{\mu}\chi_ {\nu}^*$ or $\chi_{\nu} \chi_{\mu}^*$ as unorthogonalized basis functions for the CB. This modification has no effect for the standard CD, but simplifies the discussion in case of differentiated two-electron integrals. The ambiguity in choosing a CV $\chi_{\mu} \chi_{\nu}^*$ or $\chi_\nu \chi_\mu^*$ had been already noted in Ref.~\onlinecite{Gauss23} for the two-electron integrals over GIAOs needed for computations on molecules in finite external magnetic fields.

For differentiated two-electron integrals (such as those required for the computation of nuclear forces or magnetic properties when using GIAOs), CD is not applicable, as the corresponding integrals do not form a semi-positive definite matrix. Expressions that can be used for those integrals in the CD context, however, can be obtained either via differentiation of Eq.~(\ref{eq1}) or (\ref{eq5}) or by considering CD as a special case of DF. Differentiation of Eq.~(\ref{eq1}) with respect to a perturbation $\alpha$ yields
\begin{eqnarray}
\label{eq6}
\frac{\partial (\mu \nu |\sigma \rho)}{\partial \alpha} & \approx & \sum_{P\in {\cal B}}  \left(\frac{\partial L_{\mu \nu}^P}{\partial \alpha} {L_{\rho \sigma}^P}^* + L_{\mu\nu}^P \frac{\partial {L_{\rho \sigma}^P}^*}{\partial \alpha}\right).
\end{eqnarray}
Expressions for the differentiated CVs can be obtained by taking the appropriate derivatives of the equations used for the determination of the undifferentiated CVs (see, for example, References \onlinecite{Feng19} or \onlinecite{Burger21} for a detailed discussion).

The DF-like approach on the other hand just requires to insert the auxiliary basis  (in our case the determined CB) in the expression for the derivative integrals:
\begin{eqnarray}
\label{eq7}
\frac{\partial (\mu \nu | \sigma \rho)}{\partial \alpha} = \sum_{P \in {\cal B}} \left(\left(\frac{\partial \mu \nu }{\partial \alpha}| P\right) (P | \sigma \rho) + ( \mu \nu | P) \left(P | \frac{\partial\sigma \rho}{\partial \alpha}\right) \right).
\end{eqnarray}
However,  in this way one ignores the dependence of the auxiliary basis on the perturbation. Nevertheless, nuclear forces in quantum-chemical schemes using DF are usually computed considering the geometry-dependence of the auxiliary basis,\cite{Weigend97} as this turned out be significantly more accurate.
It is easily shown that for $\tau  \rightarrow 0$ both approaches yield identical results, as the expression of the derivative approach can be rewritten as 
\begin{eqnarray}
\label{eq9}
\begin{split}
\frac{\partial (\mu \nu | \sigma \rho)}{\partial \alpha} & = \sum_{P \in {\cal B}} & \left( (\frac{\partial \mu \nu }{\partial \alpha}| P) (P | \sigma \rho) + ( \mu \nu | P) (P | \frac{\partial\sigma \rho}{\partial \alpha})  \right.  \\
&& \left. \ \ + (\mu \nu | \frac{\partial P}{\partial \alpha}) (P |\sigma \rho) + (\mu \nu | P)(\frac{\partial P}{\partial \alpha}| \sigma \rho)\right) .
\end{split}
\end{eqnarray}
The perturbed Cholesky  basis functions  can be expanded in the following manner
\begin{eqnarray}
\label{eq10}
\frac{\partial |P)}{\partial \alpha} = \sum_{P' \in {\cal B}} c_{P P'} |P')  +  \sum_{ P''\in {\cal R} \backslash {\cal B}} c_{P P''} |P'') + \sum_{P''' \in {\cal S} \backslash {\cal R}} c_{P P'''} |P''')
\end{eqnarray}
with expansion coefficients $c_{PP'}$, $c_{PP''}$, and $c_{PP'''}$.
This yields for the derivative integral
\begin{equation} \label{eq11}
\begin{split}
\frac{\partial (\mu \nu | \sigma \rho)}{\partial \alpha} &=  \sum_{P \in {\cal B}}  \left(\left(\frac{\partial \mu \nu }{\partial \alpha} \bigg| P \right) (P | \sigma \rho) + ( \mu \nu | P)\left(P \bigg| \frac{\partial\sigma \rho}{\partial \alpha} \right)\right) \\
 & + \sum_{P' \in {\cal B}} c_{PP'}  ((\mu \nu | P' ) (P |\sigma \rho) + (\mu \nu | P)(P'| \sigma \rho)) \\
  & + \sum_{P'' \in {\cal R} \backslash {\cal B}} c_{PP''}  ((\mu \nu | P'' ) (P |\sigma \rho) + (\mu \nu | P)(P''| \sigma \rho)) \\
&  + \sum_{P''' \in {\cal S} \backslash {\cal P} } c_{PP'''}  ((\mu \nu | P''' ) (P |\sigma \rho) + (\mu \nu | P)(P'''| \sigma \rho)) .
\end{split}
\end{equation}
Analysis of Eq.~(\ref{eq11}) shows that the terms in the fourth line of Eq.~(\ref{eq11}) vanish, as $(\mu \nu | P''')$ is always zero and that the contributions in the second line of Eq.~({\ref{eq11}) vanish for $\tau \rightarrow 0 $. The remaining contributions in the third 
line can be shown to cancel due to the orthonormality requirement for the perturbed Cholesky basis functions and, thus, the equivalence of the derivative and DF approach holds in the limit $\tau \rightarrow 0 $.
However, for usual choices of Cholesky thresholds in the range of $10^{-4}$ to $10^{-7} $, the equivalence between the derivative and the DF approaches does not hold.\cite{Oswald24} It is at least for geometrical-derivative integrals more accurate to include the contribution due to the perturbation dependence of the Cholesky basis functions. This is also what is standard in CD-based analytic-gradient schemes.\cite{Feng19,Schnack22,Oswald24} The situation, however, is different for derivatives with respect to an external magnetic field, as here it can be shown that the contributions due to the perturbation dependence of the Cholesky basis functions vanish.\cite{Burger21,Gauss23}

To obtain a better understanding of the CD in case of derivatives with respect to a magnetic field, it is useful to analyze the permutational symmetry of the differentiated integrals (a similar analysis does not lead to insights in the case of geometrical derivatives, as here the derivative integrals always exhibit the same permutational symmetry as the undifferentiated integrals). The differentiated integrals (with $B_i$ denoting a component of the external magnetic field) possess only fourfold permutational symmetry, i.e.,
\begin{eqnarray}
\label{eq12}
\frac{\partial (\mu \nu | \sigma \rho)}{\partial B_i} = \frac{\partial (\sigma \rho | \mu \nu)}{\partial B_i} = - \frac{\partial (\nu \mu | \rho\sigma )}{\partial B_i} = - \frac{\partial (\rho\sigma | \nu \mu)}{\partial B_i}.
\end{eqnarray}
We note in particular that the permutational symmetry with respect to the interchange of the two AO indices of electron 1 (and in the same way for electron 2) is lost. However, the latter can be still exploited in the CD of these integrals by splitting the derivative integrals into two parts
\begin{eqnarray}
\label{eq13}
\frac{\partial (\mu \nu | \sigma \rho)}{\partial B_i} = \left(\frac{\partial \mu \nu }{\partial B_i} | \sigma \rho\right) + \left( \mu \nu | \frac{\partial \sigma \rho}{\partial B_i}\right).
\end{eqnarray}
The first term on the right hand side of Eq.~(\ref{eq13}) is antisymmetric with respect to an interchange of the AOs for electron 1 and symmetric with respect to an interchange of the AOs for electron 2, while the second is symmetric for the interchange for electron 1 and antisymmetric for the interchange in case of electron 2.

Comparison of Eq.~(\ref{eq13}) with Eq.~({\ref{eq6}) shows that the differentiated CVs in the case of the magnetic field have to be antisymmetric with respect to an exchange of the two involved AOs, as the first term of Eq.~(\ref{eq6}) can only correspond to the first term on the right hand side of Eq.~(\ref{eq13}) and the second term of the right hand side Eq.~(\ref{eq6}) only to the second term on the right hand side of Eq.~(\ref{eq13}):
\begin{eqnarray}
\frac{\partial L_{\mu \nu}}{\partial B_i} = - \frac{\partial L_{\nu \mu}}{\partial B_i}.
\end{eqnarray}
We further note that the use of a symmetrized Cholesky basis functions provides a straightforward proof that there is no contribution due to the perturbation-dependence of the Cholesky basis functions, as the latter actually vanish.

For the second derivatives of the two-electron integrals with respect to an external magnetic field, the situation is more complicated. First of all, these integrals again have only fourfold permutational symmetry, i.e.,
\begin{eqnarray}
\label{eq14}
\frac{\partial^2 (\mu \nu | \sigma \rho)}{\partial B_i \partial B_j} = \frac{\partial^2 (\sigma \rho | \mu \nu)}{\partial B_i \partial B_j } = \frac{\partial^2 (\nu \mu | \rho\sigma )}{\partial B_i\partial B_j} =  \frac{\partial^2 (\rho\sigma | \nu \mu)}{\partial B_i\partial B_j},
\end{eqnarray}
but the permutational symmetry with respect to an interchange of the AOs for electron 1 and 2 can again be exploited when we split the integrals into four parts in the following way :
\begin{eqnarray}
\label{eq15}
\frac{\partial^2 (\mu \nu | \sigma \rho)}{\partial B_i\partial B_j} = \left(\frac{\partial^2 \mu \nu }{\partial B_i\partial B_j} | \sigma \rho\right) + \left( \mu\nu | \frac{\partial^2 \sigma \rho}{\partial B_i\partial B_j}\right)
+ \left(\frac{\partial \mu \nu }{\partial B_i} | \frac{\partial \sigma \rho }{\partial B_j}\right) + \left(\frac{\partial \mu \nu}{\partial B_j} | \frac{\partial \sigma \rho}{\partial B_i}\right).
\end{eqnarray}
For the first two terms on the right hand side of Eq.~(\ref{eq15}), we have symmetry with respect to an interchange of the AOs for electron 1 and 2, while we have antisymmetry for the third and fourth term.
A CD expression can be obtained for these second-derivative integrals by differentiating Eq.~(\ref{eq1}) twice with respect to the components of the magnetic field. This yields
 \begin{eqnarray}
\label{eq16}
\frac{\partial^2 (\mu \nu |\sigma \rho)}{\partial B_i \partial B_j} & \approx & \sum_{P\in {\cal B}}  \left(\frac{\partial^2 L_{\mu \nu}^P}{\partial B_i \partial B_j} {L_{\rho \sigma}^P}^* + L_{\mu\nu}^P \frac{\partial^2 {L_{\rho \sigma}^P}^*}{\partial B_i \partial B_j}
+ \frac{\partial L_{\mu \nu}^P}{\partial B_i}  \frac{\partial {L_{\rho \sigma}^P}^*}{\partial B_j} +  \frac{\partial L_{\mu\nu}^P}{ \partial B_j} \frac{\partial {L_{\rho \sigma}^P}^*}{\partial B_i}\right).
\end{eqnarray}
Based on the permutational symmetry of the various terms (dictated by the symmetry of the unperturbed and perturbed CVs), one can conclude that the terms in Eq.~(\ref{eq16}) that contain products of perturbed CVs have to correspond to the terms in Eq.~({\ref{eq15}) with derivatives in both the bra and ket, while the terms involving the doubly differentiated CVs correspond to the terms where either only the bra or the ket side is differentiated twice. We thus can make the following assignment:
\begin{eqnarray}
\label{eq17}
\sum_{P \in {\cal B}} \frac{\partial^2 L_{\mu \nu}^P}{\partial B_i \partial B_j} {L_{\rho \sigma}^P}^* \  \Rightarrow \ \left(\frac{\partial^2 \mu \nu }{\partial B_i\partial B_j} | \sigma \rho\right)  \\
\sum_{P \in {\cal B}} L_{\mu\nu}^P \frac{\partial^2 {L_{\rho \sigma}^P}^*}{\partial B_i \partial B_j} \ \Rightarrow\   \left( \mu\nu | \frac{\partial^2 \sigma \rho}{\partial B_i\partial B_j}\right) \\
\sum_{P \in {\cal B}} \frac{\partial L_{\mu \nu}^P}{\partial B_i}  \frac{\partial {L_{\rho \sigma}^P}^*}{\partial B_j} \ \Rightarrow\  \left(\frac{\partial \mu \nu }{\partial B_i} | \frac{\partial \sigma \rho }{\partial B_j}\right)  \\
\sum_{P \in {\cal B}} \frac{\partial L_{\mu\nu}^P}{ \partial B_j} \frac{\partial {L_{\rho \sigma}^P}^*}{\partial B_i} \ \Rightarrow\  \left(\frac{\partial \mu \nu}{\partial B_j} | \frac{\partial \sigma \rho}{\partial B_i}\right).
\end{eqnarray}
These relationships also provide expressions for the differentiated CVs:
\begin{eqnarray}
\label{eq18}
\frac{\partial L_{\mu \nu}^P}{\partial B_i}  &=& \left( \frac{\partial \mu \nu}{\partial B_i}| P\right) \\
\frac{\partial^2 L_{\mu \nu}^P}{\partial B_i\partial B_j}  &=& \left( \frac{\partial^2 \mu \nu}{\partial B_i\partial B_j}| P) + (\mu \nu | \frac{\partial^2 P}{\partial B_i\partial B_j}\right)
\end{eqnarray}
where we used the fact that there is no contribution from the singly differentiated Cholesky basis functions. We also note that the doubly differentiated CVs have to be symmetric with respect to an index change
\begin{eqnarray}
\frac{\partial^2 L_{\mu \nu}}{\partial B_i\partial B_j} =  \frac{\partial^2 L_{\nu \mu}}{\partial B_i\partial B_j}.
\end{eqnarray}
Within the DF picture, the second-derivative integrals are exactly represented by
\begin{eqnarray}
\begin{split}
\frac{\partial^2 (\mu \nu |\sigma \rho)}{\partial B_i \partial B_j} & = & \sum_{P\in {\cal R}}  \left( \left(\frac{\partial^2 \mu \nu}{\partial B_i \partial B_j}|P\right)(P| \sigma \rho) + (\mu \nu |P) \left(P|\frac{\partial^2 \sigma \rho}{\partial B_i \partial B_j}\right)\right.   \\ &&
+ \left. \left(\frac{\partial \mu \nu}{\partial B_i} |P\right)\left(P| \frac{\partial \sigma \rho}{\partial B_j}\right) + \left(\frac{\partial \mu \nu}{\partial B_j} |P\right)\left(P| \frac{\partial \sigma \rho}{\partial B_i}\right)\right) \
\end{split}
\end{eqnarray}
with the sum running over the basis defined by the full set of AO pairs, ${\cal R}$. Decomposing the complete basis $\cal S$ as 
\begin{eqnarray}
{\cal S} = {\cal B} + {\cal R}\backslash {\cal B} + {\cal S}\backslash {\cal R}
\end{eqnarray}
leads for the terms with the doubly differentiated AO pairs to
\begin{eqnarray}
\begin{split}
\left(\frac{\partial^2 \mu \nu }{\partial B_i\partial B_j} | \sigma \rho\right) 
= &&  \sum_{P\in {\cal B}}  \left(\frac{\partial^2 \mu \nu}{\partial B_i \partial B_j}|P\right)\left(P| \sigma \rho\right)  + \sum_{P\in {\cal R}\backslash {\cal B} } \left(\frac{\partial^2 \mu \nu}{\partial B_i \partial B_j}|P\right)\left(P| \sigma \rho\right)   \\ && + \sum_{P\in {\cal S}\backslash {\cal R}}  \left(\frac{\partial^2 \mu \nu}{\partial B_i \partial B_j}|P\right)\left(P| \sigma \rho\right) . 
\end{split}
\end{eqnarray}
The last term vanishes, as $ (P |\sigma \rho)$ is zero for $P \in {\cal S} \backslash {\cal R}$. The second term goes to zero for $\tau \rightarrow 0$ and the
first term is what one obtains from a simple resolution of the identity using the unperturbed CB. The perturbation dependence of the Cholesky basis functions
in the case of  a finite $\tau$ is thus accounted for by the second term and the question is here whether one should consider this dependence or not. One would expect that these contributions are small, but this has to be confirmed by numerical investigations.
However, considering only the first term, which the DF-like approach to the derivative integrals suggests, is certainly exact in the limit $\tau \rightarrow 0$.

For the terms with differentiated AO pairs in the bra and ket, one obtains in a similar manner 
\begin{eqnarray}
\begin{split}
 \left(\frac{\partial\mu \nu }{\partial B_i} | \frac{\partial \sigma \rho}{\partial B_j}\right) 
= &&  \sum_{P\in {\cal B}}  \left(\frac{\partial \mu \nu}{\partial B_i}|P\right)\left(P|\frac{\partial \sigma \rho}{\partial B_j}\right)  + \sum_{P\in {\cal R}\backslash {\cal B} } \left(\frac{\partial \mu \nu}{\partial B_i}|P\right)\left(P| \frac{\partial\sigma \rho}{\partial B_j}\right)   \\ && + \sum_{P\in {\cal S}\backslash {\cal R}}  \left(\frac{\partial \mu \nu}{\partial B_i}|P\right)\left(P| \frac{\partial\sigma \rho}{\partial B_j}\right) .
\end{split}\end{eqnarray}
The first term again represents the DF contribution as obtained with the unperturbed CB. The second term should again vanish for $\tau \rightarrow 0$, but no conclusion can be drawn for the third term. This term does not necessarily vanish unlike for the integrals with double differentiation in the bra or ket. The reason is simply that we cannot make any statement about the magnitude of $ (\frac{\partial \mu \nu}{\partial B_i} |P)$ when $|P) \in {\cal S} \backslash {\cal R}$. Again, numerical investigations are needed to obtain information about the magnitude of the resulting error in the integrals.

Based on this analysis, we suggest three different strategies for the treatment of the second-derivative integrals needed for the computation of the magnetizability tensor: 
\begin{enumerate}
\item{use the DF-like approach in which the perturbation-dependence of the Cholesky basis function is ignored completely. This should yield accurate results for tight Cholesky thresholds; its accuracy for loose thresholds needs to be checked.}
\item{the accuracy for the integrals involving doubly differentiated AO pairs can be improved by accounting for the perturbation dependence of the Cholesky basis functions. This can be done by computing the doubly differentiated CVs via expressions obtained by differentiation of Eq.~(\ref{eq1}).}
\item{to tackle the potential accuracy problem with the cross-derivative terms, i.e., those with derivatives in both the bra and ket, one can consider an alternative strategy for the computation of these terms. Instead of constructing them from the singly differentiated CVs, one might decompose these integral derivatives directly in a CD procedure. This is possible because these integrals constitute a Hermitian semi-positive definite matrix. The advantage of this idea which, however, comes with an increase in computational cost, is that it allows to represent this contribution to the second-derivative integrals with arbitrary accuracy.}
\end{enumerate}
The corresponding expressions for the second-derivative integrals and differentiated CVs are then  as follows:
\begin{enumerate}
\item{for the first scheme, the integrals are given by
\begin{eqnarray}
 \left(\frac{\partial^2 \mu \nu }{\partial B_i\partial B_j} | \sigma \rho\right) +  \left( \mu\nu | \frac{\partial^2 \sigma \rho}{\partial B_i\partial B_j}\right) &=&
\sum_P  \left( \frac{\partial^2 L_{\mu \nu}^P}{\partial B_i \partial B_j} {L_{\rho \sigma}^P}^*  + L_{\mu\nu}^P \frac{\partial^2 {L_{\rho \sigma}^P}^*}{\partial B_i \partial B_j}\right) \\ 
\left(\frac{\partial \mu \nu }{\partial B_i} | \frac{\partial \sigma \rho }{\partial B_j}\right)  + \left(\frac{\partial \mu \nu}{\partial B_j} | \frac{\partial \sigma \rho}{\partial B_i}\right) &=&
\sum_P  \left(\frac{\partial L_{\mu \nu}^P}{\partial B_i}  \frac{\partial {L_{\rho \sigma}^P}^*}{\partial B_j}  +\frac{\partial L_{\mu\nu}^P}{ \partial B_j} \frac{\partial {L_{\rho \sigma}^P}^*}{\partial B_i}\right)
\end{eqnarray}
and the CVs as
\begin{eqnarray}
L_{\mu \nu}^P &=&  \widetilde{(\sigma \rho | \rho \sigma)}^{-1/2}\lbrace (\mu \nu | \rho \sigma) -\sum_R^{P-1} {L_{\mu \nu}^R L_{\sigma\rho}^R}^*\rbrace   \\
\frac{\partial L_{\mu \nu}^P}{\partial B_i} &=& \widetilde{(\sigma \rho | \rho \sigma)}^{-1/2}\left\lbrace \left(\frac{\partial \mu \nu}{\partial B_i} | \rho \sigma\right) -\sum_R^{P-1} {\frac{\partial L_{\mu \nu}^R}{\partial B_i} {L_{\sigma\rho}^R}^*}\right\rbrace  \\
\frac{\partial^2 L_{\mu \nu}^P}{\partial B_i\partial B_j} &=& \widetilde{(\sigma \rho | \rho \sigma)}^{-1/2}\left\lbrace \left(\frac{\partial^2 \mu \nu}{\partial B_i\partial B_j} | \rho \sigma\right) -\sum_R^{P-1} {\frac{\partial^2 L_{\mu \nu}^R}{\partial B_i\partial B_j} {L_{\sigma\rho}^R}^*}\right\rbrace.
\end{eqnarray}
The Cholesky index $P$ is here associated with the two AO indices $\sigma$ and $\rho$ and the updated diagonal $  \widetilde{(\sigma \rho | \rho \sigma)}$ is defined as
\begin{eqnarray}
 \widetilde{(\sigma \rho | \rho \sigma)} = (\sigma \rho | \rho \sigma) - \sum_R^{P-1} {L_{\sigma \rho}^R L_{\sigma\rho}^R}^*.
\end{eqnarray}
Note also that the given expressions for the differentiated CVs can be easily implemented within the two-step procedure\cite{Aquilante11,Folkestad19,Zhang21,Gauss23} to determine the CVs (for a discussion, see Ref.~\onlinecite{Gauss23}) and that the evaluation of the differentiated CVs only requires partial integral derivatives.
}
\item{the second scheme works as the first scheme  except that the procedure for the doubly differentiated CVs is modified in order to take into account the perturbation dependence of the Cholesky basis functions. The expression for the doubly differentiated CVs is now
\begin{eqnarray}
\begin{split}
\frac{\partial^2 L_{\mu \nu}^P}{\partial B_i\partial B_j} &=& \widetilde{(\sigma \rho | \rho \sigma)}^{-\frac{1}{2}}\left\{ \left(\frac{\partial^2 \mu \nu}{\partial B_i\partial B_j} \bigg| \rho \sigma \right) + \left( \mu \nu \bigg| \frac{\partial^2\rho \sigma}{\partial B_i\partial B_j} \right)   \right. \\  &&\qquad \left. { -\sum_R^{P-1} 
\frac{\partial^2 L_{\mu \nu}^R}{\partial B_i\partial B_j} {L_{\sigma\rho}^R}^*} + \sum_R^{P-1} L_{\mu \nu}^R \frac{\partial^2 {L_{\sigma\rho}^R}^*}{\partial B_i\partial B_j}\right\}
\nonumber \\ &&
- \frac{1}{2} \widetilde{(\sigma \rho | \rho \sigma)}^{-\frac{3}{2}} \left\{
\frac{\left(\partial^2 (\sigma \rho | \rho \sigma \right)}{\partial B_i\partial B_j} \right.  \\  &&\qquad \left. { -\sum_R^{P-1} 
\frac{\partial^2 L_{\sigma \rho}^R}{\partial B_i\partial B_j} {L_{\sigma\rho}^R}^*} + \sum_R^{P-1}  L_{\sigma \rho}^R \frac{\partial^2 {L_{\sigma\rho}^R}^*}{\partial B_i\partial B_j}\right\}  \\. && \qquad\qquad
\left\{  (\mu \nu | \rho \sigma ) -\sum_R^{P-1} {L_{\mu \nu}^R L_{\sigma\rho}^R}^*\ \right\}.
\end{split}
\end{eqnarray}
}
\item{In the third scheme, the doubly differentiated CVs are obtained as in the first scheme (or the second scheme if desired), but a new set of perturbed CVs is obtained from the decomposition of the $(\frac{\partial \mu \nu}{\partial B_i }|\frac{\partial \sigma \rho}{\partial B_i}) $ integrals, where all three magnetic-field components are considered simultaneously.
The new perturbed Cholesky vectors are given by
\begin{eqnarray}
{^i}M_{\mu \nu}^P &=&
\widetilde{\left(\frac{\partial \sigma \rho}{\partial B_k} \bigg| \frac{\partial \rho \sigma}{\partial B_k}\right)}^{-\frac{1}{2}}
\left\{
 \left(\frac{\partial \mu \nu}{\partial B_i} \bigg| \frac{\partial \rho \sigma}{\partial B_k}\right)
 -\sum_R^{P-1} {^i}M_{\mu \nu}^R {^k}{M_{\sigma\rho}^R}^*
\right\}
\end{eqnarray}
The index $k$ points to the magnetic-field component that correspond to the updated diagonal $\widetilde{(\frac{\partial \sigma \rho}{\partial B_k} | \frac{\partial \rho \sigma}{\partial B_k})}$ chosen in the pivoting procedure for the actual CV, while the index $i$ denoted the magnetic-field component to which this CV contributes. The cross terms in the second-derivative integrals are then represented as
\begin{eqnarray}
\left(\frac{\partial \mu \nu }{\partial B_i} \bigg| \frac{\partial \sigma \rho }{\partial B_j}\right)  + \left(\frac{\partial \mu \nu}{\partial B_j} \bigg| \frac{\partial \sigma \rho}{\partial B_i} \right) &=&
\sum_P \left({^i}M_{\mu \nu}^P {^j}{M_{\rho \sigma}^P}^*  +{^j}M_{\mu\nu}^P {^i}{M_{\rho \sigma}^P}^*\right).
\end{eqnarray}
}
\end{enumerate}
The  choice for actual computations of the magnetizability tensor should be based on conclusions drawn from test calculations (see section~\ref{results}), as it is difficult to infer the best procedure from purely theoretical arguments. 
In the end, what matters is the accuracy of the magnetizability tensor rather than that of the second-derivative two-electron integrals and the corresponding necessary computational cost.


The use of CD in the computation of the magnetizability tensor is very similar to the use of CD in NMR chemical-shift computations.\cite{Burger21} All terms that involve unperturbed and perturbed two-electron integrals
are rewritten using the CD expressions. These terms are then evaluated directly from the unperturbed and/or perturbed CVs without a reconstruction of the two-electron integrals. For Hartree-Fock (HF) calculations of the
magnetizability tensor, the self-consistent field (SCF) and coupled-perturbed HF (CPHF)\cite{Stevens63,Pople79} steps are carried out using CVs in the same way as for the computation of nuclear magnetic shielding constants.
The only new term in the case of the magnetizability tensor is the expectation value of the doubly differentiated two-electron integrals
\begin{eqnarray}
\sum_{\mu, \nu, \sigma, \rho} \Gamma_{\mu \nu \sigma \rho} \frac{\partial^2 (\mu \nu | \sigma \rho)}{\partial B_i \partial B_j} \rightarrow \xi_{ij}
\end{eqnarray}
with $\Gamma_{\mu \nu \sigma\rho}$ as the two-particle density matrix. In the case of restricted HF(RHF) calculations, the two-particle density matrix is completely determined by the one-particle density matrix $D_{\mu \nu}$:
\begin{eqnarray}
 \Gamma_{\mu \nu \sigma \rho} =  D_{\mu \nu} D_{\sigma \rho} -  \frac{1}{2} D_{\mu \rho} D_{\sigma \nu}.
\end{eqnarray}
The two-electron contribution to the magnetizability is then given as
\begin{eqnarray}
\label{eq40}
2 \sum_P  (\sum_{\mu, \nu}  D_{\mu \nu}  L_{\mu \nu}^P ) (\sum_{\sigma, \rho} D_{\sigma \rho} \frac{\partial^2 {L^P_{\sigma \rho}}^*}{\partial B_i \partial B_j}) \nonumber \\
- \sum_P \sum_{\mu, \nu, \sigma, \rho} D_{\mu \rho} D_{\sigma \nu} L_{\mu \nu}^P \frac{\partial^2 {L^P_{\sigma \rho}}^*}{\partial B_i \partial B_j} \nonumber \\
- \sum_P \sum_{\mu, \nu, \sigma, \rho} D_{\mu \rho} D_{\sigma \nu} \frac{\partial L_{\mu \nu}^P}{\partial B_i} \frac{\partial  {L^P_{\sigma \rho}}^*}{\partial B_j}.
\end{eqnarray}
In case of scheme 3, the singly differentiated CVs are replaced by the corresponding $^iM^P$ vectors. The actual evaluation of the term in Eq.~(\ref{eq40}) is then carried out in the same way as 
for the two-electron contribution for geometrical gradients\cite{Feng19} via intermediates.

\section{Implementation}
The three suggested CD schemes for the second-derivative integrals required for the computation of the magnetizability tensor have been implemented within a development version of the \cfour program package.\cite{cfour}  The required two-electron integral derivatives are provided by means of the \mint integral package\cite{mint} and are computed using the McMurchie-Davidson scheme\cite{McMurchie78} in 
the same way as in earlier implementations for the computation of magnetizabilities using GIAO.\cite{Ruud93,Gauss07}  
Using the available unperturbed as well as singly and doubly differentiated CVs, routines for the computation of magnetizabilities at the HF level have been also added to our local \cfour version and
validated against our existing implementation not using CD.\cite{Gauss07}
 
\section{Results and Discussion}\label{results}

To compare and judge the performance of the three proposed CD treatments for the doubly differentiated two-electron integrals suggested in section~\ref{sec:theory}
calculations have been performed for several test systems. Those comprise HeH$^+$, BH, CH$_4$, H$_2$O, and vinyl alcohol.  The computations have been performed at the HF level using the cc-pVDZ and cc-pVTZ basis from Dunning's hierarchy of correlation-consistent basis sets.\cite{Dunning89} The geometries used were $r(\mathrm{HeH})$= 1.0 \AA\ for HeH$^+$, $r(\mathrm{BH})$=1.0 \AA\ for BH, $r(\mathrm{CH})$=1.00 \AA\ for CH$_4$, and $r(\mathrm{OH})$=1.0 \AA\ as well as $\langle(\mathrm{ HOH})$=104.0$^\circ$ for H$_2$O, while the geometry for vinyl alcohol was taken from Ref.~\onlinecite{Burger21}.
For vinyl alcohol, we further investigate the dependence of the accuracy on the chosen CD threshold $\tau$. We also analyze the accuracy in the computed magnetizability tensors for the three CD schemes. To demonstrate the applicability of the suggested CD schemes for the treatment of larger systems, we also report results for coronene. In the calculation for coronene, the geometry and the $\rm tz2p$ basis from Ref.~\onlinecite{Burger21} were used.

Table~\ref{table1} reports the maximum errors in the reconstructed integrals. We show the errors for three of the six different integral derivative types and also report the errors separately for 
$\lbrace (\frac{\partial^2 \mu \nu}{\partial B_i \partial B_j} |\sigma \rho) + (\mu \nu | \frac{\partial^2 \sigma \rho}{\partial B_i \partial B_j})\rbrace$ (denoted as  integrals of type $A$) and 
$\lbrace(\frac{\partial \mu \nu}{\partial B_i} |\frac{\partial\sigma \rho}{\partial B_j}) + (\frac{\partial\mu \nu}{\partial B_j} | \frac{\partial \sigma \rho}{\partial B_i})\rbrace$ (denoted as integrals of type  $B$). 

\begin{table}
\centering
 \begin{tabular}{lcccccccccc}
\hline\vspace{-0.3cm}\\
 Molecule	  &	$N_{\mathrm {AO}}$	 & $N_{\mathrm {CV}}/N_{\mathrm {CV}}^{\mathrm {total}}$	      && \multicolumn{3}{c}{type $A$} &&  \multicolumn{3}{c}{type $B$}\vspace{0.3cm}\\
&&&&  $xx$ & $yy$ & $zz$ && $xx$ & $yy$ & $zz$. \\						                                               						
\hline\\
\multicolumn{6}{l}{a) scheme 1}\vspace{0.3cm}\\
HeH$^+$      &	28	 &166/406     &&   3.7$\cdot$10$^{-5}$   & 3.7$\cdot$10$^{-5}$	&  0.0	  &&    1.9$\cdot$10$^{-3}$	& 1.9$\cdot$10$^{-3}$	&	0.0	\\
BH	          	&44	& 268/990     &&  1.1$\cdot$10$^{-4}$   &1.1$\cdot$10$^{-4}$	 & 	0.0	  &&   6.3$\cdot$10$^{-4}$	        &	6.3$\cdot$10$^{-4}$	&	0.0	\\
CH$_4$	    &     86	 &489/3471   &&  5.3$\cdot$10$^{-4 }$   & 5.2$\cdot$10$^{-4}$	&	2.6$\cdot$10$^{-4}$	  &&    6.2$\cdot$10$^{-4}$	&6.1$\cdot$10$^{-4}$	&	3.0$\cdot$10$^{-4}$	\\
H$_2$O	      &   58	& 366/1711   &&   2.5$\cdot$10$^{-4 }$  & 9.5$\cdot$10$^{-5}$	&     1.6$\cdot$10$^{-4}$	  &&  7.4$\cdot$10$^{-4}$	        &	1.7$\cdot$10$^{-4}$	&	2.1$\cdot$10$^{-3}$	\\
Vinyl alcohol & 146	& 911/10731 &&2.710$^{-4}$	       & 4.7$\cdot$10$^{-4}$	& 	3.7$\cdot$10$^{-3}$	  &&  4.5$\cdot$10$^{-4}$	        &	8.5$\cdot$10$^{-4}$	&	3.9$\cdot$10$^{-3}$	\vspace{0.3cm}\\
\multicolumn{6}{l}{b) scheme 2}\vspace{0.3cm}\\
HeH$^+$		&28	&166/406 	&&   3.2$\cdot$10$^{-5}$	 & 3.2$\cdot$10$^{-5}$	&	0.0	 &&&&\\
BH	          	&44	&268/990	   && 1.1$\cdot$10$^{-4}$	&	1.1$\cdot$10$^{-4}$	&	0.0	&&&&\\
CH$_4$	          	&86	&489/3471	 &&  5.3$\cdot$10$^{-4}$	&	5.2$\cdot$10$^{-4}$	&	2.6$\cdot$10$^{-4}$	&& \multicolumn{3}{c}{as for scheme 1}\\
H$_2$O	          	&58	&366/1711	&&   2.7$\cdot$10$^{-4}$	&	9.7$\cdot$10$^{-5}$ &	1.6$\cdot$10$^{-4}$	&&&&\\
Vinyl alcohol  &146	&911/10731  &&2.7$\cdot$10$^{-4}$	& 4.7$\cdot$10$^{-4}$	&	3.7$\cdot$10$^{-3}$	&&&& \vspace{0.3cm}\\
\multicolumn{6}{l}{c) scheme 3}\vspace{0.3cm}\\
HeH$^+$ &		28	&168/406	    &&  &&&& 8.5$\cdot$10$^{-6}$ &	8.0$\cdot$10$^{-7}$	&	0.0 \\
BH	   &       	44	& 282/990	      &&  &&&&9.0$\cdot$10$^{-6}$&		8.9$\cdot$10$^{-6}$	&	0.0 \\
CH$_4$ &	          	86	&1208/3741 &&  \multicolumn{3}{c}{as for scheme 1}&&9.7$\cdot$10$^{-6}$	&	9.7$\cdot$10$^{-6}$	&9.6$\cdot$10$^{-6}$ \\
H$_2$O	 &          	58	&699/1711	&&    &&&&   9.9$\cdot$10$^{-6}$	&	8.3$\cdot$10$^{-6}$ &	9.5$\cdot$10$^{-6}$ \\
Vinyl alcohol &    146	&2080/10731 &&   &&&&9.5$\cdot$10$^{-6	}$	&9.6$\cdot$10$^{-6}$	&	9.9$\cdot$10$^{-6}$ \\
\hline
\multicolumn{10}{l}{$^a$ for scheme 3, $N_{\mathrm {CV}}$ denoted the number of CVs obtained in the second CD.}
\end{tabular}
\caption{Maximum absolute error of the reconstructed second-derivative integrals for the three CD schemes. Reported are the errors for the integrals
$\partial^2 (\mu \nu |\sigma \rho)/ \partial B_x \partial B_x$, $\partial^2 (\mu \nu |\sigma \rho)/ \partial B_y\partial B_y$, and $\partial^2 (\mu \nu |\sigma \rho)/ \partial B_z
\partial B_z$ separately for the type $A$ and $B$ contributions. $N_{\mathrm {AO}}$ denotes the number of AOs, $N_{\mathrm {CV}}$ the number of CVs used in the calculation,$^a$ and $N_{\mathrm {CV}}^{\mathrm {total}}$ the theoretical maximum number of CVs.
All calculations were carried out with the cc-pVTZ basis and a Cholesky threshold $\tau$ of $10^{-5}$.$^a$}
\label{table1}
\end{table}																
From the results shown in Table~\ref{table1} we see that all CD schemes seem to perform more or less equally well with maximum absolute errors in the reconstructed integrals of the order of $10^{-3}$ to $10^{-6}$. This finding is in line with previous observations that for differentiated integrals the Cholesky threshold $\tau$ no longer provides a rigorous upper bound.\cite{Burger21} When comparing the three CD schemes, we see that the results for the integrals of type $A$ are  improved slightly by considering the magnetic-field dependence of the Cholesky basis.
The errors are for these integrals slightly smaller for scheme 2 in comparison with scheme 1, but this is only seen, if we analyze instead of the maximum absolute errors the standard deviation of the reconstructed integrals from the original ones. At a first glance, one might be surprised that the effect is rather small (unlike for the geometrical derivative integrals, where it is mandatory to include the perturbation dependence), but this finding can be explained: the reason for the rather small effect is that a large fraction of the Cholesky basis functions  does not depend on the magnetic field, as the two AOs of the corresponding AO pair have the same center and thus the phase factors due to the magnetic field cancel. Concerning the integrals of type $B$, we note that scheme 3 improves their accuracy significantly. In scheme 3, the Cholesky threshold again provides an upper bound for the error in the reconstructed integrals and indeed the reported error is in all cases smaller than $10^{-5}$.

Table~\ref{table2} reports the computed magnetizability tensor for our test systems. However, as there are no visible differences for the results obtained with the three different CD schemes, we only show one set of values. We further note that these values agree to all reported digits with corresponding numbers from computations without using CD.
\begin{table}{}
\centering
 \begin{tabular}{lccccccc}
\hline
 Molecule &  $\xi_{xx}$ & $\xi_{xy}$ & $\xi_{xz}$ & $\xi_{yy}$ & $\xi_{yz}$ & $\xi_{zz}$ & $\xi$ \\
 \hline
HeH$^+$	&	-0.412	&0.000	&0.000	&-0.412	&0.000	&-0.354	&-0.393	\\
BH		&	7.912	&0.000	&0.000	&7.912	&0.000	&-2.402	&4.474	\\
CH$_4$	        &-3.816	&0.000	&0.000	&-3.816	&0.000	&-3.816	&-3.816	\\
H$_2$O		&	-2.937	&0.000	&0.000	&-2.843&	0.000	&-2.905	&-2.895	\\
Vinyl alcohol &    -6.248&	-0.057	&0.000	&-5.148	&0-000	&-6.676	&-6.024	\\
\hline\\
\end{tabular}
\caption{Computed magnetizability tensor (at HF/cc-pVTZ level, in a.u.) for the test systems. 
With a Cholesky threshold of $\tau$ of $10^{-5}$, computations with all three schemes yield the same values within the reported digits.
$\xi_{ij} $ denotes the individual components of the magnetizability tensor and $\xi$ the isotropic magnetizability
defined as one third of the trace of the magnetizability tensor.}
\label{table2}
\end{table}

In Table~\ref{table3} we show the error in the reconstructed integral derivatives for vinyl alcohol (cc-pVDZ basis) as a function of the Cholesky threshold $\tau$.
\begin{table}{}
\centering
 \begin{tabular}{lccccccccc}
\hline\vspace{-0.3cm}\\
 $\tau$ & $N_{\mathrm {CV}}/N_{\mathrm {CV}}^{\mathrm {total}}$ && \multicolumn{3}{c}{type $A$} &&  \multicolumn{3}{c}{type $B$}\vspace{0.3cm}\\
&&& $xx$ & $yy$ & $zz$ && $xx$ & $yy$ & $zz$ \\
 \hline\vspace{-0.3cm}\\
\multicolumn{6}{l}{a) scheme 1}\vspace{0.3cm}\\													
	$10^{-4}$ &	305/1953	&&1.3$\cdot$10$^{-3}$	&2.1$\cdot$10$^{-3}$	&	9.9$\cdot$10$^{-3}$	&&4.0$\cdot$10$^{-3}$ &	8.9$\cdot$10$^{-3}$	&1.0$\cdot$10$^{-2}$	\\
	$10^{-5}$ &400/1953	&&5.3$\cdot$10$^{-4}$	&3.5$\cdot$10$^{-4}$	&	1.2$\cdot$10$^{-3}$	&&8.8$\cdot$10$^{-4}$ &	2.6$\cdot$10$^{-3}$	&1.7$\cdot$10$^{-3}$	\\
	$10^{-6}$ &	503/1953	&&4.8$\cdot$10$^{-5}$	&4.7$\cdot$10$^{-5}$	&	2,9$\cdot$10$^{-4}$	&&3.6$\cdot$10$^{-4}$ &	1.2$\cdot$10$^{-3}$	&	4.5$\cdot$10$^{-4}$	\\
	$10^{-7}$ &640/1953	&&6.5$\cdot$10$^{-6}$	&	8.2$\cdot$10$^{-6}$	&	2.5$\cdot$10$^{-5}$	&&1.1$\cdot$10$^{-4}$ &	4.7$\cdot$10$^{-4}$	&	1.2$\cdot$10$^{-4}$	\\
	$10^{-8}$ &796/1953	&&1.8$\cdot$10$^{-6}$	&	2.9$\cdot$10$^{-6}$	&	3.9$\cdot$10$^{-6}$	&&5.8$\cdot$10$^{-5}$ &	2.7$\cdot$10$^{-4}$	&	3.8$\cdot$10$^{-5}$	\\
	$10^{-9}$ &937/1953	&&2.7$\cdot$10$^{-7}$	&	4.2$\cdot$10$^{-7}$	&	3.2$\cdot$10$^{-7}$	&&2.5$\cdot$10$^{-5}$ &	1.5$\cdot$10$^{-4}$	&	2.4$\cdot$10$^{-5}$	\vspace{0.3cm}\\
\multicolumn{6}{l}{b) scheme 2}\vspace{0.3cm}\\
	$10^{-4}$ &305/1953	&&1.2$\cdot$10$^{-3}$	&	2.1$\cdot$10$^{-3}$	&	9.8$\cdot$10$^{-3}$\\
	$10^{-5}$ &400/1953	&&4.4$\cdot$10$^{-4}$	&3.2$\cdot$10$^{-4}$	&	1.3$\cdot$10$^{-3}$\\
	$10^{-6}$ &503/1953	&&4.4$\cdot$10$^{-5}$	&	2.7$\cdot$10$^{-5}$	&	2.5$\cdot$10$^{-4}$ &&  \multicolumn{3}{c}{as for scheme 1}\\
	$10^{-7}$ &640/1953	&&5.6$\cdot$10$^{-6}$	&	4.1$\cdot$10$^{-6}$	&	2.1$\cdot$10$^{-5}$\\
	$10^{-8}$ &796/1953	&&5.6$\cdot$10$^{-7}$	&	6.1$\cdot$10$^{-7}$	&	3.3$\cdot$10$^{-6}$\\
	$10^{-9}$ &937/1953	&&6.2$\cdot$10$^{-8}$	&	2.1$\cdot$10$^{-7}$	&	1.2$\cdot$10$^{-7}$\vspace{0.3cm}\\
\multicolumn{6}{l}{c) scheme 3}\vspace{0.3cm}\\								
$10^{-4}$	&679/1953	&&&&&&9.0$\cdot$10$^{-5}$	&	9.0$\cdot$10$^{-5}$	&	9.7$\cdot$10$^{-5}$	\\
$10^{-5}$	&942/1953	&&&&&&9.9$\cdot$10$^{-6}$	&	9.8$\cdot$10$^{-6}$	&	9.5$\cdot$10$^{-6}$	\\
$10^{-6}$	&1284/1953	&& \multicolumn{3}{c}{as for scheme 1}&&1.0$\cdot$10$^{-6}$	&	9.8$\cdot$10$^{-7}$	&	9.5$\cdot$10$^{-7}$	\\
$10^{-7}$	&1646/1953	&&&&&&9.0$\cdot$10$^{-8}$	&	9.9$\cdot$10$^{-8}$	&	9.0$\cdot$10$^{-8}$	\\
$10^{-8}$	&1953/1953	&&&&&&2.9$\cdot$10$^{-8}$	&	2.8$\cdot$10$^{-8}$	&	2.7$\cdot$10$^{-8}$	\\
$10^{-9}$	&1953/1953	&&&&&&1.1$\cdot$10$^{-7}$	&	1.2$\cdot$10$^{-7}$	&	9.7$\cdot$10$^{-8}$	\\
\hline
\multicolumn{10}{l}{$^a$ for scheme 3, $N_{\mathrm {CV}}$ denotes the number of CVs obtained in the second CD.}
\end{tabular}
\caption{Maximum absolute error of the reconstructed second-derivative integrals for the three CD schemes for different Cholesky thresholds $\tau$. Reported are the errors for integrals
$\partial^2 (\mu \nu |\sigma \rho)/ \partial B_x \partial B_x$, $\partial^2 (\mu \nu |\sigma \rho)/ \partial B_y\partial B_y$, and $\partial^2 (\mu \nu |\sigma \rho)/ \partial B_z
\partial B_z$ separately for the type $A$ and $B$ contributions. $N_{\mathrm {CV}}$ denoted the number of CVs used in the calculation$^a$ and $N_{\mathrm {CV}}^{\mathrm {total}}$ the theoretical maximum number of CVs.
All calculations have been performed with the cc-pVDZ basis.}
\label{table3}
\end{table}						
The findings from Table~\ref{table3} are also visualized in Figure~\ref{figure1} where the maximum error for the $xx$ integrals is plotted as a function of the Cholesky threshold.
\begin{figure}[h!]
    \center
    \includegraphics[width=0.9\textwidth]{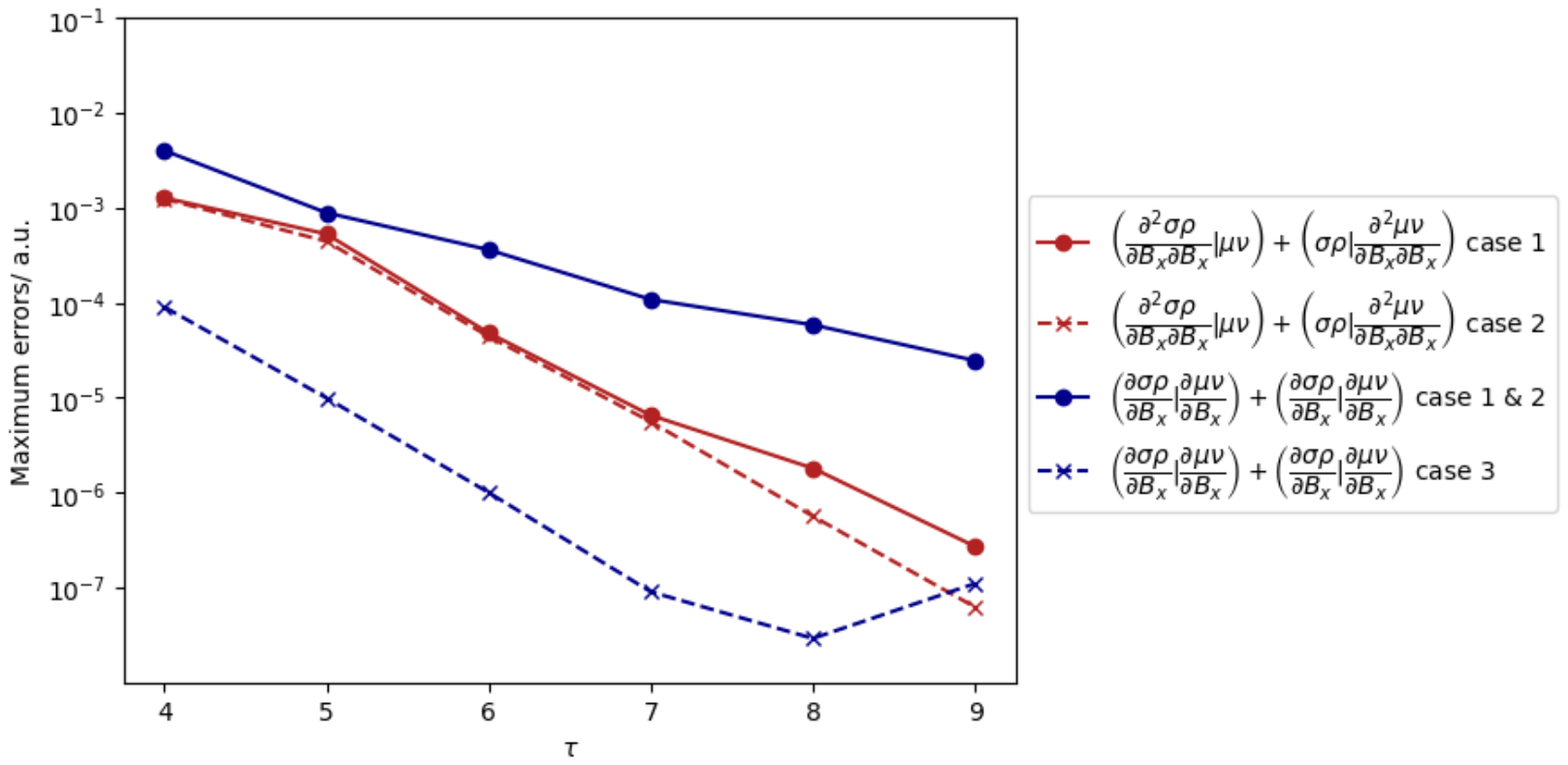}
    \caption{Maximum absolute error of the reconstructed $\partial^2 (\mu \nu |\sigma\rho)/\partial B_x \partial B_x$ integrals as a function of the used Cholesky threshold $\tau$ for vinyl alcohol and the cc-pVDZ basis.}
    \label{figure1}
\end{figure}
The error in the reconstructed two-electron integrals of type $A$ vanishes for $\tau \rightarrow 0$, as it should be based on theoretical arguments. The differences between scheme 1 and 2
are not visible for loose Cholesky thresholds and only show up for tight thresholds. This is not unexpected, as the number of magnetic-field dependent Cholesky basis functions increases 
with a tighter threshold. Thus, 68~$\%$ of the Cholesky basis functions have the same center for $\tau = 10^{-4}$ (see Table~\ref{table4}), but the number drops to 48~$\%$ for $\tau = 10^{-6}$ and 32~$\%$  for $\tau = 10^{-9}$. 
For the integrals of type $B$, we note that in the case of scheme 1 and 2, the error does not go to zero but instead settles at values around $10^{-5}$. This is what is expected based on the incompleteness of the set ${\cal R}$, as discussed section~\ref{sec:theory}. 
We note that the remaining error is small (and does not seriously affect the computed magnetizability tensors of our test systems). Nevertheless, the residual errors in the integrals of type $B$  seen for scheme 1 and 2 can be eliminated by means of scheme 3, as here the error of the reconstructed integrals vanishes for $\tau \rightarrow 0$. The computed magnetizabilities show no variations with $\tau$ in the reported digits and all computed numbers agree with those given in Table~\ref{table2}.
\begin{table}{}
\centering
\begin{tabular}{lcc}
\hline
$\tau$ && $N_{\mathrm {CV}}^{\mathrm {same\ center}}$ \\
\hline
$10^{-4}$ && 30 (68 \%)\\
$10^{-5}$ && 30 (57 \%)\\
$10^{-6}$ && 30 (48 \%)\\
$10^{-7}$ && 30 (39 \%)\\
$10^{-8}$ && 30 (37 \%)\\
$10^{-9}$ && 30 (32 \%)\\
\hline
\end{tabular}
\caption{Number of CVs that originate from pairs of AOs located at the same center as a function of the Cholesky threshold $\tau$. The percentage of CVs that originate from such AO pairs is given in parentheses. }
\label{table4}
\end{table}						

Finally, we report results for the magnetizability tensor of coronene (C$_{24}$H$_{12}$) as obtained at the HF level using a tz2p basis (684 AOs). Figure~\ref{figure2} shows the structure of the coronene molecule. Table~\ref{table5} 
reports the computed magnetizability tensors.
\begin{figure}[h!]
    \center
    \includegraphics[width=0.9\textwidth]{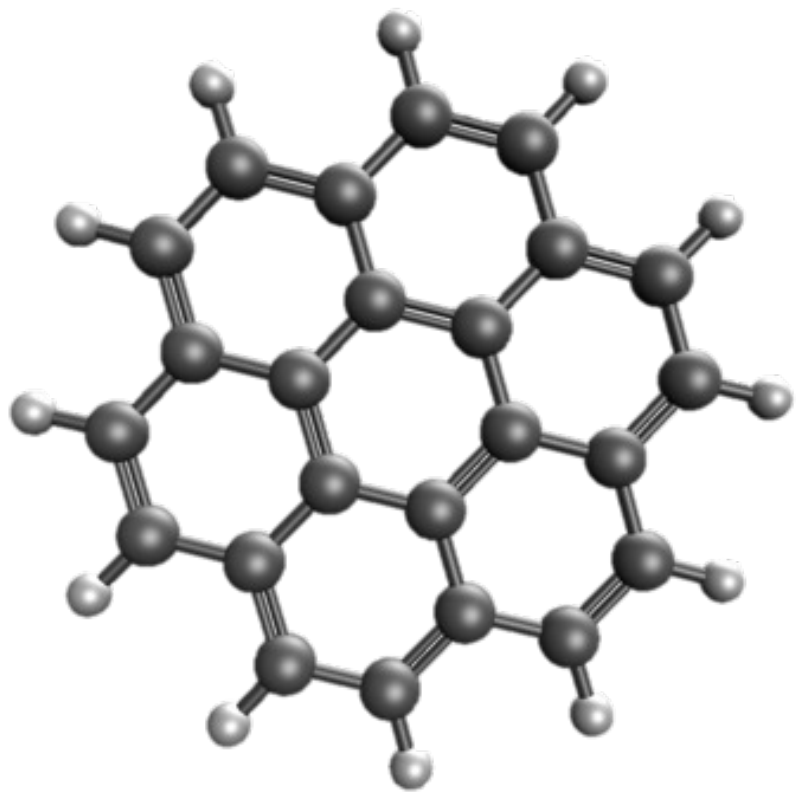}
    \caption{Structure of the coronene molecule (C$_{24}$H$_{12}$).}
    \label{figure2}
\end{figure}
\small
\begin{table}{}
\centering
\begin{tabular}{lccccccccc}
\hline
CD treatment  & $N_{\mathrm {CV}}$ & $\xi_{xx}$ &$\xi_{yy}$ &$\xi_{zz}$ &$\xi$ & $t(\mathrm {CD})$  & $t(\mathrm {total})$ \\
\hline
scheme 1&		5473	    &  -25.083 & -25.088&-130.301 & -60.157 &  14583 & 18471	\\
scheme 2&		5473    &  -25.083 & -25.088 &	-130.301 & -60.157& 26213 & 30103	\\
scheme 3 & 5473+15746	    & -25.082  & -25.087 & -130.300 & -60.157& 14675+81840	& 86027	\\
\hline 
\multicolumn{10}{l}{$^a$ for scheme 3, $N_{\mathrm {CV}}$ and $t({\mathrm {CD}})$ are given for the integrals of type $A$ and $B$ separately.}
\end{tabular}
\caption{Computed magnetizability tensor (in a.u.) for coronene (HF/tz2p computations with $\tau = 10^{-6}$). Reported are the values for the non-vanishing individual components, $\xi_{ij}$, the isotropic
magnetizability, $\xi$, the number of CVs, $N_{\mathrm {CV}}$, and the times (in s) needed to perform the CD as well as the total computation. Calculations have been performed with 8 cores on a Intel(R) Xeon(R) CPU E5-2643 v3 @ 3.40GHz node.$^a$}
\label{table5}
\end{table}						
\normalsize
First of all, all schemes yield for a threshold $\tau = 10^{-6}$ the same results. With respect to the timings we note that scheme 1 is the fastest 
requiring only a fraction of the time of scheme 3. This is expected, as scheme 1 and 2 do not require the computation of the cross terms (integrals of type $B$) and scheme 1 only requires the computation of 
partial second derivatives with derivative contributions only at the right hand side.

Based on the results of our test computations and the computations for coronene, we recommend scheme 1 for the routine computations of magnetizability tensors. 
The scheme provides accurate results, in comparison to computation without use of CD,  and is computationally the most efficient.
´
\section{Conclusion}

In this paper, we propose three different CD schemes for the handling of the doubly differentiated two-electron integrals required in the computation of the magnetizability tensor when using GIAOs.
The first scheme is based on the DF interpretation of the CD of the two-electron integrals and just inserts the Cholesky basis in the doubly differentiated two-electron integrals. Scheme 2 is based
on the differentiation of the CD expression for the unperturbed two-electron integrals. The third scheme addresses the issue that the first two schemes are not able to represent the cross terms
in the doubly differentiated integrals with arbitrary accuracy. It is suggested to decompose the cross terms which constitute a Hermitian semi-positive matrix directly in a CD procedure and use
the resulting CVs instead of the perturbed CVs of scheme 1 and 2.

The performance of the three proposed schemes is judged by means of calculations for HeH$^+$, BH, CH$_4$, H$_2$O, and vinyl alcohol. While some, though modest differences  are seen
in the accuracy of the integrals reconstructed from the CVs, no visible difference  is noted in the computed HF values for the magnetizability tensor elements.
This finding is confirmed in large-scale computations for the magnetizability tensor of coronene (C$_{24}$H$_{12}$) with a tz2p basis (684 AOs). We thus recommend the use
of scheme 1 for the routine computation of magnetizability tensors in the CD context.

While we have performed so far only HF computations of the magnetizability tensor using CD decomposed two-electron integrals, future work will focus on the implementation of CD schemes
for the computation of magnetizability tensors at MP2 and the complete active-space self-consistent field (CASSCF) level, thereby exploiting existing codes for the computation of nuclear magnetic shielding constants.\cite{Burger21,Nottoli22}

\begin{acknowledgement}
This paper is dedicated to Trygve Helgaker on the occasion of his 70th birthday in 2023. JG thanks him for almost 30 years of friendship, his great
hospitality during many visits to Oslo (including a sabbatical
and a stay at the Centre for Advanced Study at the Norwegian
Academy of Science and Letters), and many exciting scientific
collaborations. SS thanks him for his support, mentorship, and his inspiring enthusiasm about molecules in magnetic fields which sparked her own interest in the field. Both, SS and JG thank him also for a first and very enjoyable introduction into cross-country skiing in 2013.
The authors acknowledge funding by the Deutsche Forschungsgemeinschaft (DFG) within
the project B5 of the TRR 146 (Project No. 233630050).
\end{acknowledgement}






\bibliography{main}

\end{document}